\documentclass[11pt]{article}
\fontfamily{times}

\usepackage{geometry, amsmath}
\usepackage{bm}
\usepackage{bbm}
\usepackage{epsf}
\usepackage{epsfig}
\usepackage{xcolor}
\usepackage{listings}
\usepackage{amsthm}

\usepackage{soul}
\usepackage[style=authoryear,sorting=ynt,natbib=true,bibstyle=numeric]{biblatex}
\renewbibmacro{in:}{}
\usepackage{graphicx}
\usepackage{float}
\usepackage{caption}
\usepackage{subcaption}
\usepackage{tikz}
\usetikzlibrary{decorations.markings}
\usetikzlibrary{arrows.meta}
\newtheorem{thm}{Theorem}

\newtheorem{rem}{Remark}
\newtheorem{lemma}{Lemma}

\newcommand{\ben}{\begin{enumerate}}
\newcommand{\een}{\end{enumerate}}
\newcommand{\beq}{\begin{eqnarray}}
\newcommand{\eeq}{\end{eqnarray}}
\newcommand{\beqn}{\begin{eqnarray*}}
\newcommand{\eeqn}{\end{eqnarray*}}

\newcommand{\N}{{\cal{N}}}

\newcommand{\be}{\begin{equation}}
\newcommand{\ee}{\end{equation}}

\def\sk1{\vskip 10pt}

\def\b{{\bold b}}

\def\b0{{\bold 0}}

\def\bold{\bf}

\providecommand{\keywords}[1]{\textbf{\textit{Keywords: }} #1}

\def\ben{\textcolor{red}}

\newenvironment{prof}[1][Proof]{\noindent\textit{#1}\quad }
{\hfill $\Box$\vspace{0.7mm}}
 
\usepackage{amssymb} 
\renewcommand{\Bbb}{\mathbb}

\title{Early Detection of Treatment’s Side Effect: A Sequential Approach}
\author {}
\author{Jiayue Wang\thanks{
Email: jwa9@iu.edu } \ and  Ben Boukai\thanks{
Email: bboukai@iupui.edu}  \\ Department of Mathematical Sciences, Indiana University Indianapolis\\
Indianapolis, Indiana, 46202 }

\begin{document}
\maketitle


\begin{abstract}

\noindent With the emergence and spread of infectious diseases with pandemic potential, such as COVID-19, the urgency for vaccine development have led to unprecedented compressed and accelerated schedules that shortened the standard development timeline. In a relatively short time, the leading pharmaceutical companies\footnote{e.g. Pfizer--BioNTech, Moderna}, received an Emergency Use Authorization (EUA) for vaccine's en-mass deployment. To monitor the potential side effect(s) of the vaccine during the (initial) vaccination campaign, we developed an optimal sequential test that allows for the early detection of potential side effect(s). This test employs a rule to stop the vaccination process once the observed number of side effect incidents exceeds a certain (pre-determined) threshold. The optimality of the proposed sequential test is justified when compared with the $(\alpha, \beta)$ optimality of the non-randomized fixed-sample Uniformly Most Powerful (UMP) test. 

In the case of a single side effect, we study the properties of the sequential test and derive the exact expressions of the Average Sample Number (ASN) curve of the stopping time (and its variance) via the regularized incomplete beta function. Additionally, we derive the asymptotic distribution of the relative 'savings' in ASN as compared to maximal sample size. Moreover, we construct the post-test parameter estimate and studied its sampling properties, including its asymptotic behavior under local-type alternatives. These limiting behavior results are the consistency and asymptotic normality of the post-test parameter estimator. We conclude the paper with a small simulation study illustrating the asymptotic performance of the point and interval estimation and provide a detailed example, based on COVID-19 side effect data (see \citet{beatty2021analysis}) of our suggested testing procedure.

\end{abstract}



\bigskip
\vfill\eject
\setlength{\parindent}{0pt}

\section{Introduction} \label{s1}

The emergence and spread of infectious diseases with pandemic potential, such as influenza, cholera etc., occurred regularly throughout history. The most recent example is the one caused the human Coronavirus, (SARS CoV-2), dubbed as COVID-19, which was first identified in human in the late 2019. Among the multifaceted public health responses (isolation, shutdown, masking, etc.), development of a vaccine and vaccination of the population appeared to be the most promising one for slowing the spread of the virus and preventing millions of deaths. Usually, vaccines development takes several years, as it must go through a number of phases of clinical trials to test its safety, immunogenicity, effectiveness, dose levels and possible adverse effects. However, the urgency to create a vaccine for COVID‑19 led to unprecedented compressed and accelerated schedules that shortened the standard timeline for vaccine development. In less than a year of development, the two leading pharmaceutical companies Pfizer–BioNTech and Moderna received on December 11, 2020, an Emergency Use Authorization (EUA) for their (mRNA based) vaccines. Almost immediately, in January 2021, the deployment of the vaccination campaign started en masse through vaccination centers across the country. However, due to rapid timeline of development and the mRNA base, the need for monitoring potential side effects of the vaccine became more acute. Therefore, after the individuals vaccinated, they were all required to stay on site to be monitored for at least $15$ minutes in case they exhibited some side effects (fainting, allergic reaction, dizziness, nausea, etc.). Presumably, once there were recorded too many cases of side effect(s), the vaccination process would have been stopped (at least for that day and vaccination center). 

Based on such a procedure for monitoring people after being vaccinated, we develop a formal sequential procedure for an early detection of such side effect(s). In particular, we are interested in the formal sequential testing of the hypotheses
\begin{equation} \label{1.1}
H_0: \theta \le \theta_0 \ \ \text{against} \ \ H_1: \theta > \theta_0,
\end{equation}
based on a sequence of independent binary responses, where $\theta$ denotes the probability of exhibiting a side effect to the vaccine ($0<\theta<1$) and $\theta_0$ denotes the maximal tolerable probability for such side effect to the new vaccine (very small, say, $\theta_0<0.1$). Accordingly, the vaccination process should be stopped once $H_0$ is rejected, (that is if it deemed $\theta>\theta_0$); otherwise, the vaccination process should be continued. Clearly, we aim to continue the vaccination campaign as long as $H_0: \theta \le \theta_0$ is not rejected.

However, note that \citet{COVID-19} (FDA) conducted sequential safety test for anaphylaxis, utilizing Poisson Maximized Sequential Probability Ratio Test (PMaxSPRT) and Binomial Maximized Sequential Probability Ratio Test (BMaxSPRT) of the relative risk (the rate of each safety outcome compared to a control baseline rate). Moreover, they also used the disproportionality analysis i.e. proportional reporting ratio (PRR) in the further signal detection of adverse events. However, here we use a more direct approach of sequential hypotheses testing in \eqref{1.1} to test for the proportion (the probability of potential side effect(s) compared to a nominal threshold).

In the next section, we introduce the classical non-randomized fixed-sample Uniformly Most Powerful (UMP) test for binomial proportion which achieves the desired optimality for given Type~I error probability and Type~II error probability, $(\alpha, \beta)$. In section \ref{ss2.1}, we develop a corresponding optimal sequential test for a curtailed sampling scheme, that detects and stops the sampling (vaccination) as early as possible and also matches the desired optimality of the UMP test. As was already discussed in \citet{eisenberg1980curtailed}, the correspondence between the optimality of the curtailed randomized sequential test and that of the UMP test is well-known in literature. However, in the case of our proposed sequential test, we construct and justify the optimality criterion as compared with the non-randomized UMP test. Here, in order to study the properties of our proposed sequential test, we use the notable relationship between cumulative (negative) binomial probabilities and regularized incomplete beta function (see \citet{deveaux2018clinical}). We exploit this relationship to study the Average Sample Number (ASN) function, and other related quantities of our proposed procedure. Based on the proposed optimal sequential test, the effective random sample size, upon termination, follows a negative binomial distribution. This fact was utilized by \citet{patil1963equivalence} who provided the expression of ASN of a sequential procedure for binomial probabilities involving the acceptance and rejection boundaries of an acceptance sampling plan. A version of that acceptance sampling plan was studied by \citet{phatak1967estimation}, which included a version of our proposed sequential test along with the estimation of ASN via maximum likelihood estimation (MLE). In the current paper, however, we derive the ASN and the related quantities in a direct manner, not relying on the asymptotic normal distribution of the MLE.

Note that the classical Sequential Probability Ratio Test (SPRT) for binomial probabilities, which includes three situations (Rejecting, Accepting and Continuing sampling) has been modified and studied numerous times in literature, see for example \citet{armitage1957restricted}, \citet{anderson1960modification}, \citet{alling1966closed}, \citet{breslow1970sequential}, just to name a few. However, in the context of vaccination campaign and the sequential testing of the hypotheses in (\ref{1.1}), we are only concerned in the two cases: rejecting $H_0$ or not rejecting and continuing sampling (i.e. continue the vaccination campaign). We study the general properties of our proposed sequential procedure along with the sampling properties of the post-test estimator of the model parameter. This includes exact calculations of the expectation and the variance of this post-test estimator. We augment these results with the asymptotic properties of our estimator based on local-type alternatives. Specifically, we prove the consistency and asymptotic normality of the post-test estimator, which allow us to construct, upon termination, confidence interval for the unknown model parameter. Moreover, we also provide the limiting expression of relative 'savings' under our sequential testing procedure. We conclude the paper with a small simulation study demonstrating the asymptotic performance of the point and interval estimation. We also provide a detailed example based on COVID-19 side effect data (see \citet{beatty2021analysis}) illustrating our sequential testing procedure.

\section{The Sequential detection procedure} \label{s2}

As we described in the Introduction, we are interested in the early detection of the side effect, in a situation in which the observations are made sequentially, as during the vaccination process, say. To that end, we develop a sequential testing procedure that matches the optimality criteria of the classical (fixed-sample) Uniformly Most Powerful (UMP) test. To fix the idea, consider the testing problem of the hypotheses in \eqref{1.1} conducted in the fixed-sample situation. That is, draw a simple random sample of $N$ individuals from the 'underlying population' for inspection. Each tested person in the sample can be treated as a Bernoulli process that indicates whether the person exhibited the side effect with the unknown probability $\theta$. Accordingly, we denote by $S_N=X_1+X_2+\cdots+X_N$ (where $X_i\sim \mathcal{B}(1,\theta)$, $i.i.d.$), the total number of individuals with side effect observed among the $N$ vaccinated people. Note that, $ \forall \ \theta \in (0,1)$, the number of individuals with side effect observed among the $N$ people, namely, is a binomial random variable, $S_N \sim \mathcal{B}(N,\theta)$\footnote{The $pmf$ of the Binomial distribution, $\mathcal{B}(n,\theta)$, is given by $p(j \mid n, \theta)={\genfrac{(}{)}{0pt}{}{n}{j}}\theta^j(1-\theta)^{n-j}\mathbf{I}[j=0, 1, \dots, n].$}. Under the binomial assumption, the (non-randomized fixed-sample) UMP testing procedure with a Type~I error probability not exceeding a desired level $\alpha$ ($\alpha\in (0,1)$), rejects $H_0$ whenever $S_N$ is larger than some predetermined critical value $k$ (to be later specified). That is, the fixed-sample test of the hypotheses in \eqref{1.1}, denoted here as $\operatorname{T_{fix}}$, is
\begin{equation} \label{2.2}
	\operatorname{T_{fix}}:=
	\begin{cases}
		\text{if} \ S_N >k,  & \text{Reject} \ H_0;\\ 
		\text{if} \  S_N \le k,  & \text{Do not reject} \ H_0. 
	\end{cases}
\end{equation}
Denoting by $\Pi_{\operatorname{T_{fix}}}(\theta)$ the power function of this test, it follows immediately from \eqref{2.2} that 
\begin{equation} \label{2.3}
\Pi_{\operatorname{T_{fix}}}(\theta)=\operatorname{P}_{\theta}(\operatorname{T_{fix}} \text{ rejects } H_0)=\operatorname{P}_{\theta}(S_N>k), \quad  0< \theta <1, 
\end{equation}
where $P_\theta(\cdot)$ denote probability under the Binomial $pmf$ of $S_N$. We note that the binomial tail probability, $\operatorname{P}_{\theta}(S_N>k)$, has a useful integral representation (see \citet{hartley1951chart}), given  
\begin{equation} \label{2.4}
\operatorname{P}_{\theta}(S_N>k)=\mathcal{I}_\theta(k+1,N-k), 
\end{equation}
where $\mathcal{I}_\cdot(\cdot,\cdot)$ denotes the regularized incomplete beta function, namely, for any $\xi>0, a>0$ and $b>0$,
\begin{equation} \label{2.5}
\mathcal{I}_\xi(a,b):=\int_0^\xi f(u \mid a, b) du\equiv \int_0^\xi \frac{u^{a-1}(1-u)^{b-1}}{b(a, b)}du,
\end{equation}
with $f(u \mid a,b)$ being the $pdf$ of a ${\cal{B}}eta(a, b)$ random variable. We utilize this convenient representation \eqref{2.4} repeatedly in the sequel. 

It immediately follows from \eqref{2.4} and \eqref{2.5} that $\mathcal{I}_{\theta}(k+1,N-k)$ is strictly increasing in $\theta$ for any $N\geq 1$ and any $0\leq k\leq N$. Hence $\Pi_{\operatorname{T_{fix}}}(\theta)=\operatorname{P}_{\theta}(S_N>k)$ is monotonically increasing as a function of $\theta$. Accordingly, for the test $\operatorname{T_{fix}}$ to have a \emph{desired size} of at most $\alpha$, ($\alpha\in (0,1)$), we have by the monotonicity of the power function in \eqref{2.3},  
\begin{equation} \label{2.6}
\tilde{\alpha}:= \sup_{\theta\leq \theta_0} \operatorname{P}_{\theta}(\operatorname{T_F} \text{ rejects } H_0)\equiv \sup_{\theta\leq \theta_0} \operatorname{P}_{\theta}(S_N >k)=\operatorname{P}_{\theta_0}(S_N >k)\leq \alpha. 
\end{equation}
 Additionally if for a given $\theta_1>\theta_0$, we wish the Type~II error probability of $\operatorname{T_{fix}}$ not to exceed a specified desired value $\beta$, $(\beta \in (0,1))$, we have,  
\begin{equation} \label{2.7}
\tilde{\beta}(\theta_1) =1-\Pi_{\operatorname{T_{fix}}}(\theta_1)= \operatorname{P}_{\theta_1}(S_N \le k)\le \beta.
\end{equation}

Now for given $\alpha$, $\beta$, $\theta_0$ and $\theta_1$, we may simultaneously 'solve' equations \eqref{2.6} and \eqref{2.7} for $N$ and $k$ to obtain the {\it{optimal}} UMP test of $H_0$ versus $H_1$ in \eqref{1.1}. Namely, one can determine the optimal 'sample size', $N^*\equiv N(\alpha, \beta, \theta_0, \theta_1)$ and a corresponding 'critical test value', $k^*$, by either an iterative procedure utilizing \eqref{2.6} and \eqref{2.7} and the Binomial $pmf$ (see Footnote $1$) or by the standard Normal approximations to the Binomial probabilities\footnote{See conditions in $(1)-(2)$ of \citet{schader1989two}.} are given by,  
\begin{equation} \label{2.8}
N^*=\left[ \left(\frac{z_{1-\alpha}\sqrt{\theta_0(1-\theta_0)}+z_{\beta}\sqrt{\theta_1(1-\theta_1)}}{\theta_1-\theta_0}\right)^2 \right] , 
\end{equation}
and,
\begin{equation} \label{2.9}
k^*=\left[ N^*(z_{1-\alpha}\sqrt{\frac{\theta_0(1-\theta_0)}{N^*}}+\theta_0)-\frac{1}{2} \right], 
\end{equation}
where $[x]$ is the nearest integer value to $x$ and $z_p:=\Phi^{-1}(p)$, $\forall \ p \in (0,1)$ where $\Phi$ denotes the standard Normal $cdf$.

\subsection{The Optimal Sequential test} \label{ss2.1}

Now suppose that for the given desired probabilities of Type~I and Type~II errors, $(\alpha, \beta)$, and the given $\theta_0$ and $\theta_1$, the optimal values $N^*$ and $k^*$ are determined as was described above. In view of the sequential nature of the vaccination process (the sampling process), we consider sequential version of the fixed-sample test $\operatorname{T_{fix}}$ described above.  

In that framework, the data is a sequence of observations on the vaccinated people $X_1, X_2, \dots$ that become available one–at–a–time or in batches. Let $S_n=X_1+X_2+ \dots +X_n$ be the total number of individuals exhibiting the side effect among the {\it{first}} $n$ vaccinated individuals. Once $S_n$ reaches a predefined critical value $k^*+1$, the collection of data is ceased which provides the termination of the observation (vaccination) process and gives rise to a random sample size $M_{k^*}$, that is, a \emph{stopping time} (see \citet{woodroofe1982nonlinear} for details), defined in this case as, 
\be \label{2.10}
M_{k^*}=\inf\{n > k^*: S_n\geq k^*+1\}.
\ee
The sequential test of \eqref{1.1} is therefore can be written as:
\be \label{2.11}
\operatorname{T_{seq}}: =
\begin{cases}
\text{if} \ S_n=k^*+1, \  \ & \text{stop and reject $H_0$, $M_{k^*}=n$ and $S_{M_{k^*}}=k^*+1$}; \\
\text{if} \ S_ n \le k^*, \  \ & \text{continue the vaccination}. 
\end{cases}
\ee
It can be easily ascertained that since the sequential test \eqref{2.10} is based on a sequence of independent Bernoulli trials, each with a probability $\theta$ of a side effect, which is terminated once a predefined number, $k^*+1$, of side effect(s) has been observed, the stopping time $M_{k^*}$ has a negative binomial distribution, $M_{k^*} \sim \mathcal{NB}(k^*+1,\theta)$\footnote{The $pmf$ of the Negative Binomial distribution $\mathcal{NB}(r,\theta)$ is given by $p(j \mid r, \theta)={\genfrac{(}{)}{0pt}{}{j-1}{r-1}}\theta^r(1-\theta)^{j-r}\mathbf{I}[j=r, r+1, \dots ]$.}, with a mean and variance given by, 
$$
E_{\theta}(M_{k^*} )=\frac{k^*+1}{\theta}, \quad Var_{\theta}(M_{k^*} )=\frac{(k^*+1)(1-\theta)}{\theta^2}.
$$
However, since in the most realistic situations, the daily supply of vaccines available to the vaccination center is limited to $N^*$ units per day (say), the sequential observation (vaccination) process must be terminated once $N^*$ has been reached. Thus, upon termination, the sequential testing procedure involves a terminal decision rule as it is based on $(X_1, X_2, \dots, X_{M^*})$, where
\be \label{2.12}
M^*=\min \{ M_{k^*} ,N^* \} \equiv M_{k^*}  \wedge N^*.
\ee 
Note that the corresponding sequential test, now curtailed, can be written equivalently in terms of the stopping time $M_{k^*}$ as:
\be \label{2.13}
\operatorname{T^*_{seq}}: =
\begin{cases}
	\text{if} \ M_{k^*}  \le N^*, \  \ & \text{Stop and reject $H_0$;} \\
	\text{if} \ M_{k^*} >N^*, \  \ & \text{Do not reject $H_0$.} 
\end{cases}
\ee

In Figure \ref{f1}, we illustrate the sample paths for $S_n$, upon rejection (black) and upon non-rejection (brown). 

\begin{figure}[H]
	\centering
	\begin{tikzpicture}[>=Latex,scale=0.3]
		\draw[help lines, color=gray!60](0,0) grid (20,15);
		\fill (0,0) node[left]{$O$} circle (.1);
		\draw[thick,->] (0,0) -- (22,0) node[anchor=north west]{$n$};
		\draw[thick,->] (0,0) -- (0,17) node[anchor=south west]{$S_n$};
		\draw (19,0) node[below]{$N^*$} circle (.2);
		\draw (19,1)  circle (.2);
		\draw (19,2)  circle (.2);
		\draw (19,3)  circle (.2);
		\draw (19,4)  circle (.2);
		\draw (19,5)  circle (.2);
		\draw (19,6)  circle (.2);
		\draw (19,7)  circle (.2);
		\draw (19,8)  circle (.2);
		\draw (19,9) node[right]{$(N^*,k^*)$} circle (.2);
		\fill (0,10) node[left]{$k^*+1$} circle (.1);
		\fill (10,0) node[below]{$k^*+1$}circle (.1) ;
		\fill (10,10)  circle (.2);
		\fill (11,10)  circle (.2);
		\fill (12,10)  circle (.2);
		\fill (13,10)  circle (.2);
		\fill (14,10)  circle (.2);
		\fill (15,10)  circle (.2);
		\fill (16,10)  circle (.2);
		\fill (17,10)  circle (.2);
		\fill (18,10)  circle (.2);
		\fill (19,10)  circle (.2);
		\draw[semithick, dashed] (0,0) -- (10,10);
		\fill (1,1)  circle (.1);
		\fill (2,2)  circle (.1);
		\fill (3,3)  circle (.1);
		\fill (4,4)  circle (.1);
		\fill (5,5)  circle (.1);
		\fill (6,6)  circle (.1);
		\fill (7,7)  circle (.1);
		\fill (8,8)  circle (.1);
		\fill (9,9)  circle (.1);
		\draw[semithick,solid] (0,0) -- (1,0)-- (4,3)-- (5,3)-- (7,5)-- (8,5)-- (13,10);
		\fill (1,0)  circle (.1);
		\fill (2,1)  circle (.1);
		\fill (3,2)  circle (.1);
		\fill (4,3)  circle (.1);
		\fill (5,3)  circle (.1);
		\fill (6,4)  circle (.1);
		\fill (7,5)  circle (.1);
		\fill (8,5)  circle (.1);
		\fill (9,6)  circle (.1);
		\fill (10,7)  circle (.1);
		\fill (11,8)  circle (.1);
		\fill (12,9)  circle (.1);
		\draw[semithick,solid,color=brown] (0,0) -- (2,0)--(4,2)--(6,2)-- (7,3)-- (8,3)--(9,4)-- (11,4)--(13,6)--(15,6)--(16,7)--(17,7)--(19,9);
		\fill [brown] (1,0)  circle (.1);
		\fill [brown](2,0)  circle (.1);
		\fill [brown](3,1)  circle (.1);
		\fill [brown](4,2)  circle (.1);
		\fill [brown](5,2)  circle (.1);
		\fill [brown](6,2)  circle (.1);
		\fill [brown](7,3)  circle (.1);
		\fill [brown](8,3)  circle (.1);
		\fill [brown](9,4)  circle (.1);
		\fill [brown](10,4)  circle (.1);
		\fill [brown](11,4)  circle (.1);
		\fill [brown](12,5)  circle (.1);
		\fill [brown](13,6)  circle (.1);
		\fill [brown](14,6)  circle (.1);
		\fill [brown](15,6)  circle (.1);
        \fill [brown](16,7)  circle (.1);
        \fill [brown](17,7)  circle (.1);
         \fill [brown](18,8)  circle (.1);
         \fill [brown](19,9)  circle (.2);
		\fill [color=brown](20,4) node[right]{-- stop vaccination and not reject};
		\fill (20,6) node[right]{-- stop vaccination and reject};
	\end{tikzpicture}
	\caption{The curtailed sequential test $\operatorname{T^*_{seq}}$}
	\label{f1}
\end{figure}

\begin{rem} \label{rem1}
It follows immediately from \eqref{2.13} that the power function of the $\operatorname{T^*_{seq}}$ is
\be\label{2.14}
\Pi_{\operatorname{T^*_{seq}}}(\theta)=\operatorname{P}_\theta(M_{k^*} \leq N^*), \qquad \forall \theta \in (0,1).
\ee
\end{rem}
In Theorem \ref{thrm1} below we establish that $\Pi_{\operatorname{T^{}_{fix}}}(\cdot)=\Pi_{\operatorname{T^*_{seq}}}(\cdot)$, so that the curtails sequential test $\operatorname{T^*_{seq}}$ in \eqref{2.13} is also optimal in the sense of $\operatorname{T_{fix}}$ in \eqref{2.2}. 

\begin{thm}\label{thrm1}
Let $\operatorname{T_{fix}}$ be the optimal $(\alpha,\beta)$ UMP test of $H_0$ versus $H_1$ in \eqref{1.1} with fixed $N^*$ and $k^*$ and a corresponding power function $\Pi_{\operatorname{T_{fix}}}(\theta)$ as given in \eqref{2.3}. Let $\operatorname{T^*_{seq}}$, in \eqref{2.13}, be the curtailed sequential test of the hypotheses \eqref{1.1}, with a power function $\Pi_{\operatorname{T^*_{seq}}}(\theta)$, then we have, 
\be \label{2.15}
\Pi_{\operatorname{T^{}_{fix}}}(\theta)=\Pi_{\operatorname{T^*_{seq}}}(\theta), \quad \forall \ \theta \in (0,1).
\ee
Hence, for the sequential test $\operatorname{T^*_{seq}}$ it holds that, 
\be \label{2.16}
\Pr\left(\text{Type~I error}\right) \le\Pi_{\operatorname{T^*_{seq}}}(\theta_0)\leq \alpha\ \ \  \text{and}  \ \ \ \Pr\left(\text{Type~II error}\right) \le1-\Pi_{\operatorname{T^*_{seq}}}(\theta_1)\le \beta.
\ee
\end{thm}

\begin{prof} 
First note that, 
$$ 
\{M_{k^*} \leq N^*\} \equiv \bigcup_{k^*+1}^{N^*}\{S_n \geq k^*+1\}. 
$$
Moreover, since for $j=0, \dots n$, $\{S_n\geq j\}\subseteq\{S_{n+1}\geq j\}$, we have $\bigcup_{k^*+1}^{N^*}\{S_n\geq k^*+1\}=\{S_{N^*}\geq k^*+1\}\equiv \{S_{N^*}> k^*\}$, and hence
\be\label{2.17}
\operatorname{P}_\theta(S_{N^*}>k^*)\equiv \operatorname{P}_\theta(M_{k^*} \leq N^*). 
\ee
Since by \eqref{2.14}, $\Pi_{\operatorname{T^*_{seq}}}(\theta)=\operatorname{P}_\theta(M_{k^*} \leq N^*)$ and by \eqref{2.3}, $\Pi_{\operatorname{T_{fix}}}(\theta)=\operatorname{P}_{\theta}(S_{N^*}>k^*)$, we conclude from \eqref{2.17} that $\Pi_{\operatorname{T_{fix}}}(\theta)=\Pi_{\operatorname{T^*_{seq}}}(\theta)$, $\forall \ \theta \in (0,1)$.  Finally, since by \eqref{2.4} and \eqref{2.5}, $\Pi_{\operatorname{T_{fix}}}(\theta)$ is monotonically increasing with respect to $\theta$, we immediately obtain the same properties of $\Pi_{\operatorname{T^*_{seq}}}(\theta)$ in \eqref{2.16} as were obtained in \eqref{2.6} and \eqref{2.7} for $\Pi_{\operatorname{T_{fix}}}(\theta)$.
\end{prof}

Clearly by Theorem \ref{thrm1} above, our proposed $\operatorname{T^*_{seq}}$ is the \emph{optimal} sequential test, which is as powerful as the non-randomized $(\alpha,\beta)$ UMP test $\operatorname{T_{fix}}$.

 \begin{rem}\label{rem2}
While \citet{rider1962negative} established, in similarity to \eqref{2.4}, the relationship between the Negative Binomial probabilities and the regularized incomplete beta function, we note that since equation \eqref{2.17} holds for any $N>1$ and $0\leq k\leq N$, it follows immediately from \eqref{2.4} that 
\be\label{2.18}
\operatorname{P}_\theta \left(S_{N}>k \right)\equiv \operatorname{P}_\theta \left(M_k\leq N \right)=\mathcal{I}_\theta \left(k+1, N-k \right). 
\ee
Accordingly, with the optimal $N^*$ and $k^*$ at hand, it also follows from \eqref{2.17} and \eqref{2.18}
\be\label{2.19}
\Pi_{\operatorname{T^*_{seq}}}(\theta)\equiv \operatorname{P}_\theta(M_{k^*} \leq N^*)\equiv \mathcal{I}_\theta \left(k^*+1, N^*-k^* \right). 
\ee
\end{rem}

Next, we study the efficiency of our proposed sequential test. Usually, this efficiency is measured by the Average Sample Number (ASN) function, $ASN^*(\theta):= E _{\theta}\left( M^* \right)$. Note that, by its definition, $M^*=M_{k^*} \wedge N^*$, we trivially obtain
$$
E _{\theta}\left( M^* \right)\leq \min \left\{N^*, E_\theta(M_{k^*} )\right\}=\min\left\{N^*, \frac{k^*+1}{\theta}\right\}.
$$
which is less than $N^*$ at any $\theta\in (0,1)$.

\begin{thm}\label{thrm2}
For $\forall$ $\theta\in (0, 1)$, let $ASN^*(\theta)$ be the ASN function of the optimal sequential test $\operatorname{T^*_{seq}}$ defined above, then 

\be\label{2.20}
ASN^*(\theta)\equiv E_{\theta}\left( M^* \right)=N^*\mathcal{I}_{1-\theta}\left(N^*-k^*,k^*+1\right)+\frac{k^*+1}{\theta}\mathcal{I}_{\theta}\left(k^*+2,N^*-k^*\right), 
\ee
and 
\begin{multline}
 \label{2.21}
E_{\theta}\left({M^*}^2\right)={N^*}^2\mathcal{I}_{1-\theta}\left(N^*-k^*,k^*+1\right)+\frac{(k^*+1)(k^*+2)}{\theta^2}\mathcal{I}_{\theta}\left(k^*+3,N^*-k^*\right)\\
-\frac{k^*+1}{\theta}\mathcal{I}_{\theta}\left(k^*+2,N^*-k^*\right).
\end{multline}

\end{thm}

\begin{prof}
Since $M^*=M_{k^*}  \wedge N^*$, we have 
\be \label{2.22}
E _{\theta}\left( M^* \right)=N^* \operatorname{P}_{\theta}\left(M_{k^*} >N^*\right)+E_{\theta} \left(M_{k^*}  \mathbbm{1}[M_{k^*}  \le N^*] \right) 
\ee
where $\mathbbm{1}[{\cal A}]$ is the indicator function of the set. 
By utilizing \eqref{2.18}, we may rewrite the first part in \eqref{2.22} as
$$
N^* \operatorname{P}_{\theta}\left(M_{k^*} >N^*\right)=N^*\mathcal{I}_{1-\theta}\left(N^*-k^*,k^*+1\right).
$$
Since $M_{k^*} \sim \mathcal{NB}(k^*+1,\theta)$, by direct calculation, we obtain the second part in \eqref{2.22} as
$$
E_{\theta} \left(M_{k^*}  \mathbbm{1}[M_{k^*}  \le N^*] \right)=\frac{k^*+1}{\theta}\operatorname{P}_{\theta}\left(\tilde{M}\le N^*+1\right),
$$
where $\tilde{M}\sim \mathcal{NB}\left(k^*+2,\theta\right)$, which together with \eqref{2.18} (see also Remark \ref{rem2}), provides that 
\be \label{2.23}
E_{\theta} \left(M_{k^*}  \mathbbm{1}[M_{k^*}  \le N^*] \right)=\frac{k^*+1}{\theta}\mathcal{I}_{\theta}\left(k^*+2,N^*-k^*\right).
\ee
Finally, by combining them together, we prove \eqref{2.20}. To prove \eqref{2.21}, we note that 
\begin{align} \label{2.24}
E_{\theta}\left({M^*}^2\right)=&{N^*}^2\operatorname{P}_{\theta}\left(M_{k^*}>N^*\right)+E_{\theta} \left(M_{k^*}^2 \mathbbm{1}[M_{k^*} \le N^*] \right) \notag\\
=& {N^*}^2\mathcal{I}_{1-\theta}\left(N^*-k^*,k^*+1\right)+E_{\theta} \left(M_{k^*}^2 \mathbbm{1}[M_{k^*} \le N^*] \right).
\end{align}
Since $M_{k^*}\sim \mathcal{NB}(k^*+1, \theta)$ and 
\begin{align*}
E_{\theta} \left(M_{k^*}\left(M_{k^*}+1\right) \mathbbm{1}[M_{k^*} \le N^*] \right)=&
\frac{(k^*+1)(k^*+2)}{\theta^2}\operatorname{P}_\theta \left( \tilde{M} \le N^*+2 \right) \\
\\
\equiv &\frac{(k^*+1)(k^*+2)}{\theta^2}\mathcal{I}_{\theta}\left(k^*+3,N^*-k^*\right),
\end{align*}
where we utilized \eqref{2.18} with $\tilde{M} \sim \mathcal{NB}\left(k^*+3,\theta\right)$. Then applying the value of $E_{\theta}\left(M_{k^*}\mathbbm{1}[M_{k^*} \le N^*] \right)$ in \eqref{2.23}, we immediately obtain the second part in \eqref{2.24} as
\be \label{2.25}
E_{\theta} \left(M_{k^*}^2 \mathbbm{1}[M_{k^*} \le N^*] \right)=\frac{(k^*+1)(k^*+2)}{\theta^2}\mathcal{I}_{\theta}\left(k^*+3,N^*-k^*\right)-\frac{k^*+1}{\theta}\mathcal{I}_{\theta}\left(k^*+2,N^*-k^*\right).
\ee
Next, by substituting \eqref{2.25} into \eqref{2.24}, we obtain \eqref{2.21}. 
\end{prof}

Clearly, upon utilizing Theorem \ref{thrm2}, we may obtain a computable expression for $Var_{\theta}\left(M_{k^*}\right)$, by combining the value of $E_{\theta}\left(M_{k^*}\right)\equiv ASN^* (\theta)$ in \eqref{2.20} and the value of $E_{\theta}\left(M^2_{k^*}\right)$ in \eqref{2.21}.

Figure \ref{f2} below illustrates the ASN curve and the power function, $\Pi_{\operatorname{T^*_{seq}}}(\theta)$, with respect to $\theta$ ($\forall \ \theta \in (0,1)$). From the plots, we can see that once the value of $\theta$ exceeds $\theta_0$, both ASN curve and power function change rapidly. It indicates that our proposed sequential test does perform as we expected.

\begin{figure}[H]
\centering
\begin{subfigure}[b]{0.45\textwidth}
\centering
\includegraphics[width=\textwidth]{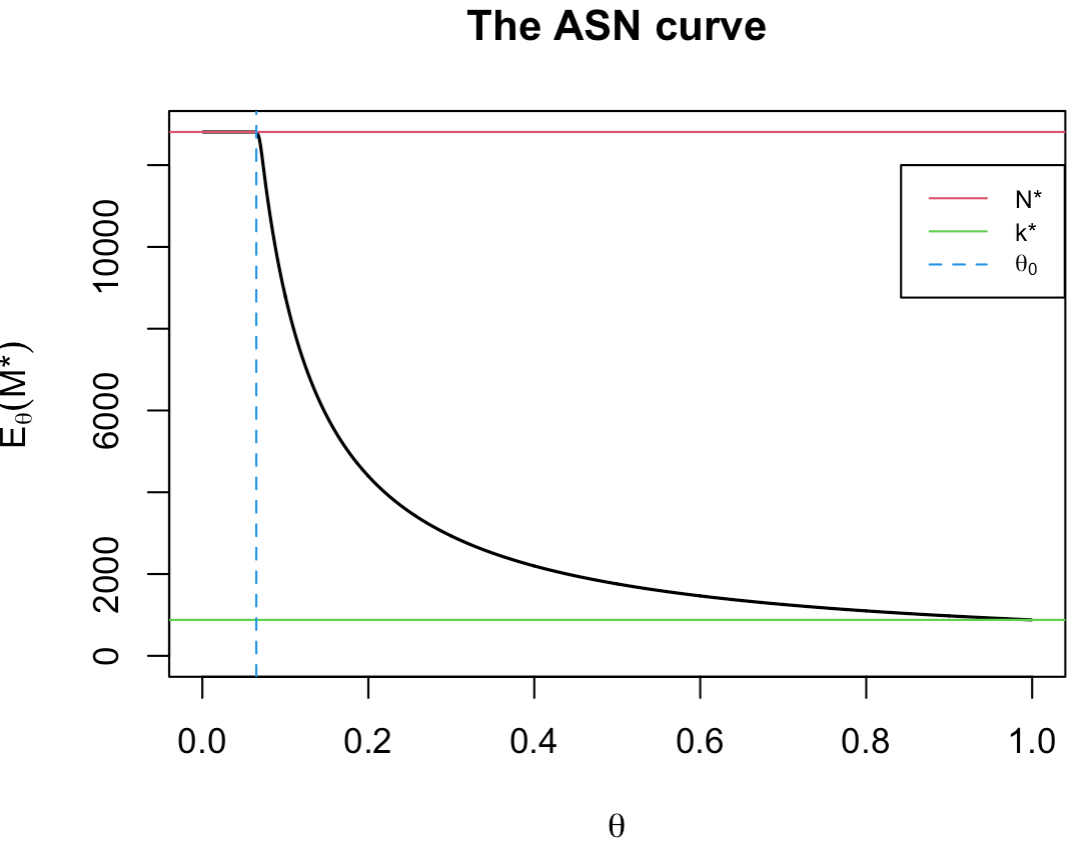}
\end{subfigure}
\hfill
\begin{subfigure}[b]{0.45\textwidth}
\centering
\includegraphics[width=\textwidth]{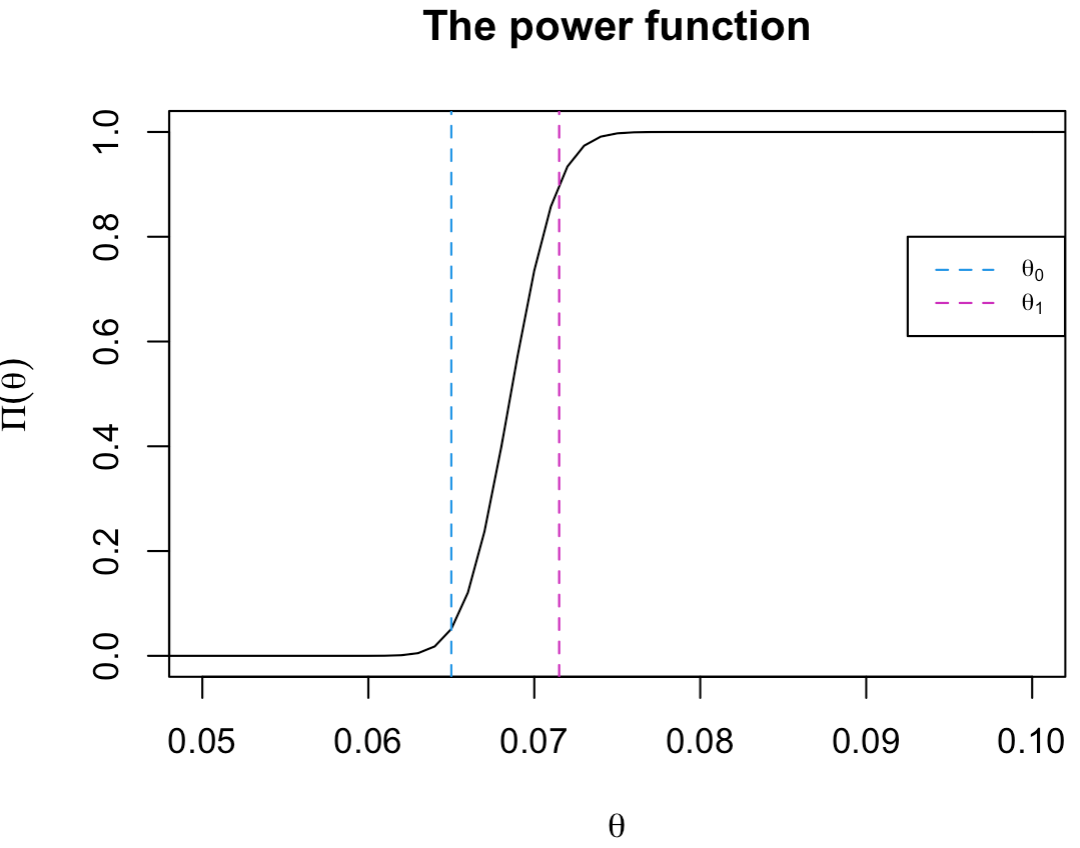}
\end{subfigure}
\caption{The plots of the $ASN$ curve and the power function with respect to $\theta$ for fixed $\alpha=0.05$, $\beta=0.1$, $\theta_0=0.065$, $\theta_1=0.0715$ with $N^*=12811$ and $k^*=878$.}
\label{f2}
\end{figure}


In the following table we present the value of the coefficient of variation (CV), intended to measure the relative dispersion of the terminal sample size, $M^*$ (with respect to $\theta$), 
$$
CV(\theta) := \frac{\sqrt{Var_\theta(M^*)}}{E_\theta(M^*)},
$$
as is calculated for a choice of the design parameter of this curtailed sequential procedure. As it appears from Table \ref{t1}, the proposed sequential test results are with relatively low CV for the range of $\theta$ values of interest.

\begin{table}[H]
\begin{center}{\small 
\caption{The expectation, the standard deviation and the CV of $M^*$ with respect to $\theta$ for fixed $\alpha=0.05$, $\beta=0.1$, $\theta_0=0.065$, $\theta_1=0.0715$ with $N^*=12811$ and $k^*=878$.}
\label{t1}
\begin{tabular}{c|c|c|c}
\hline \rule{0pt}{12pt}
$\theta$  &  $E_\theta\left({M^*}\right)$ &  $\sqrt{Var_\theta(M^*)} $  & $CV(\theta)$   \  \\ \hline
0.065	 & 12802  & 52.5240   & 0.0041        \  \\ 
0.0715	 & 12274  & 363.9850   & 0.0297        \  \\ 
0.1	 & 8790   & 281.2650	 & 0.0320    \  \\ 
0.2	  & 4395  & 132.5896  & 0.0302      \  \\ 
0.3	  & 2930  & 82.6841  & 0.0282        \  \\ 
0.4	  & 2198   & 57.4130   &  0.0261        \  \\ 
0.5	  & 1758  & 41.9285   &  0.0239        \  \\ 
\hline
\end{tabular}
}
\end{center}
\end{table}

\section{Post-Test Estimation} \label{s3}

Following the implementation of this $(\alpha,\beta)$-optimal sequential testing procedure, we are also interested, upon termination, in the estimation of the unknown rate, $\theta$, of the side effect. In this section, we study the properties of the post-test estimate of this parameter as well as those properties of the effective random sample size (at termination) in an asymptotic framework.

Clearly, for this sequential testing procedure, if upon termination, $H_0$ is not rejected, we have a fixed number of vaccinated people $N^*$ of which $S_{N^*}$ of them have exhibited the side effect. Otherwise, if the null hypothesis is rejected, the testing procedure involves a random sample size $M_{k^*}$ with $k^*+1$ of the people exhibiting the side effect. Accordingly at termination, the post-test estimator of $\theta$ is,

\be \label{3.26}
\hat{\theta}_{M^*}=\frac{S_{N^*} }{N^*} \mathbbm{1}\left[S_{N^*}\leq k^*\right] +\frac{k^*+1}{M_{k^*}}\mathbbm{1}\left[M_{k^*} \le N^*\right].
\ee
Clearly, the expected value of $\hat{\theta}_{M^*}$ is, 
\be\label{3.27}
E_{\theta}\left(\hat{\theta}_{M^*}\right)=E_{\theta}\left( \frac{S_{N^*} }{N^*}\mathbbm{1}\left[S_{N^*}\leq k^*\right] \right) +E_{\theta}\left(\frac{k^*+1}{M_{k^*}}\mathbbm{1}\left[M_{k^*} \le N^*\right] \right). 
\ee
Since $S_{N^*}\sim {\cal B}(N^*, \theta)$, direct calculation yields, the first part in \eqref{3.27} as 
\be \label{3.28}
E_{\theta}\left( \frac{S_{N^*} }{N^*}\mathbbm{1}\left[S_{N^*}\leq k^*\right] \right)=\theta\operatorname{P}_{\theta}\left(S_{N^*-1} \le k^*-1\right),
\ee
which could easily be evaluated using \eqref{2.18}. Also by direct calculation, the second part in \eqref{3.27} can be written as 
\be \label{3.29}
E_{\theta}\left(\frac{k^*+1}{M_{k^*}}\mathbbm{1}\left[M_{k^*} \le N^*\right] \right)=(k^*+1) \sum_{j=k^*+1}^{N^*}\frac{1}{j}\operatorname{P}_{\theta}(M_{k^*}=j), 
\ee
where $\operatorname{P}_{\theta}(M_{k^*}=j)$ is calculated from the $pmf$ of the $\mathcal{NB}(k^*+1, \theta)$ distribution (see Footnote $3$). Similarly, the second moment of $\hat{\theta}_{M^*}$ can be expressed as
\be \label{3.30}
E_{\theta}(\hat{\theta}^2_{M^*})=E_{\theta}\left(\frac{S_{N^*}^2 }{{N^*}^2}\mathbbm{1}\left[S_{N^*}\leq k^*\right]  \right)+E_{\theta}\left(\frac{(k^*+1)^2}{M^2}\mathbbm{1}\left[M \le N^*\right] \right).
\ee
Since $E_{\theta}\left(S_{N^*}(S_{N^*}-1)\mathbbm{1}\left[S_{N^*}\leq k^*\right]\right)=N^*(N^*-1)\theta^2 \operatorname{P}_{\theta}\left(S_{N^*-2} \le k^*-2 \right)$ and by utilizing \eqref{3.28}, we obtain that the first part in \eqref{3.30} may be written as
\be \label{3.31}
E_{\theta}\left(\frac{S_{N^*}^2 }{{N^*}^2}\mathbbm{1}\left[S_{N^*}\leq k^*\right]\right)=\frac{(N^*-1)\theta^2\operatorname{P}_{\theta}\left(S_{N^*-2}\le k^*-2\right)}{N^*}+\frac{\theta\operatorname{P}_{\theta}\left(S_{N^*-1} \le k^*-1\right)}{N^*}, 
\ee
which could also be easily evaluated numerically by \eqref{2.18}. Finally, by direct calculation (see also Footnote $4$), we obtain the second part of \eqref{3.30} as   
\be \label{3.32}
E_{\theta}\left(\frac{(k^*+1)^2}{M^2_{k^*}}\mathbbm{1}\left[M_{k^*} \le N^*\right] \right)=(k^*+1)^2 \sum_{j=k^*+1}^{N^*}\frac{1}{j^2}\operatorname{P}_{\theta}(M_{k^*}=j).
\ee
Combining \eqref{3.31} and \eqref{3.32} for the second moment, as well as \eqref{3.28} and \eqref{3.29} for the first moment of $\hat{\theta}_{M^*}$, we can immediately obtain the corresponding expression for the variance of $\hat{\theta}_{M^*}$, (which could readily be evaluated numerically utilizing \eqref{2.18} and the $pmf$ of the Negative Binomial distribution (see Footnote $4$)). 

The following table presents the calculated values of the first two moments and the variance of $\hat{\theta}_{M^*}$ under the same parameterization as in Table \ref{t1}. As can be seen, $E_{\theta}(\hat{\theta}_{M^*})$ is close to $\theta$ with the corresponding diminishing small variance.
\begin{table}[H]
\begin{center}{\small 
\caption{The first, second moments and the variance of $\hat{\theta}_{M^*}$ with respect to $\theta$ for fixed $\alpha=0.05$, $\beta=0.1$, $\theta_0=0.065$, $\theta_1=0.0715$ with $N^*=12811$ and $k^*=878$.}
\label{t2}
\begin{tabular}{c|c|c|c}
\hline
$\theta$  &  $E_\theta\left( \hat{\theta}_{M^*} \right)$ &  $E_\theta\left( \hat{\theta}_{M^*}^2 \right) $  & $Var_\theta\left( \hat{\theta}_{M^*} \right)$   \  \\ \hline
0.065	 & 0.0647  & 0.0042   & 2.0804e-05       \  \\ 
0.0715	 & 0.0712  & 0.0051   & 3.5520e-05      \  \\ 
0.1	 & 0.1001   & 0.0100	 & 1.0279e-05    \  \\ 
0.2	  & 0.2002  & 0.0401  & 3.6521e-05    \  \\ 
0.3	  & 0.3002 & 0.0902  &  7.1852e-05  \  \\ 
0.4	  & 0.4003  & 0.1603  & 1.0941e-04      \  \\ 
0.5	  & 0.5003  & 0.2504  & 1.4237e-04   \  \\ 
\hline
\end{tabular}
}
\end{center}
\end{table}

\subsection{Asymptotic Properties} \label{ss3.1}

We consider the asymptotic behavior of our sequential testing procedure under \emph{'local-type'} alternatives, as $\theta_1 \to \theta_0$, more specifically as $\theta_1=\theta_0(1+\delta)$ with $\delta \rightarrow 0$ \footnote{Alternatively, one may take $\theta_1=\theta_0+\delta$, in the subsequent derivations which lead to the same asymptotic results described here.}. Note that for given $\alpha, \ \beta, \ \theta_0$ and $\theta_1=\theta_0(1+\delta)$, one can easily determine the corresponding optimal values of $N^*$ and $k^*$ as are given in equations \eqref{2.8} and \eqref{2.9}. Specifically, with $\theta_1=\theta_0(1+\delta)$, these optimal values, $N^*_\delta \equiv N^*(\alpha, \beta, \theta_0, \delta)$ and $k^*_\delta \equiv k^*(\alpha, \beta, \theta_0, \delta)$ are given by 
\be \label{3.33}
N^*_\delta=\left[ \left(\frac{z_{\alpha}}{\delta}\sqrt{\frac{1-\theta_0}{\theta_0}}+z_{\beta}\sqrt{\frac{1+\delta}{\delta^2}\left(\frac{1}{\theta_0}-1-\delta\right)}\right)^2 \right], 
\ee
and
\be \label{3.34}
k^*_\delta=\left[ N^*_\delta \left(z_\alpha\sqrt{\frac{\theta_0(1-\theta_0)}{N^*_\delta}}+\theta_0 \right)-\frac{1}{2} \right].
\ee
In Lemma \ref{lem1} below we establish the asymptotic behaviors of $N^*_\delta$ and $k^*_\delta$ as $\delta \to 0$. 

\begin{lemma} \label{lem1}
 With the optimal $N^*_\delta$ and $k^*_\delta$ as are given in \eqref{3.33} and \eqref{3.34} above, we have, as $\delta\to 0$, $N^*_\delta \to \infty$ and $k^*_\delta \to \infty$ in such a way so that 
$$
\lim_{\delta\to 0}\frac{k^*_\delta}
{N^*_\delta} = \theta_0.
$$
\end{lemma}
The proof of this result is given in Section \ref{s6} below.

The next result describes the asymptotic behavior of the 'relative savings' in the number of observations attained by our sequential testing procedure. To simplify the subsequent notation, we denote $M_{\delta}\equiv M_{k_\delta^*}$ and $M^*_\delta\equiv \min\{M_{k^*_\delta}, N^*_\delta \}$.

\begin{thm} \label{thrm3}
As $\delta \to 0$, we have the asymptotic \ul{relative} 'savings' in the Average Sample Number as compared to $N^*_{\delta}$ is
\be \label{3.35}
E_\theta\left(\frac{N^*_{\delta}-M^*_{\delta}}{N^*_{\delta}}\right) \longrightarrow \left(1-\frac{\theta_0}{\theta}\right)^{+}, \quad\forall \ \theta \in (0,1),
\ee
where $(x)^{+}=\max\{0,x\}$.
\end{thm}

\begin{prof}
First note that by \eqref{2.20}, we have 
\be \label{3.36}
N^*_\delta -E_\theta\left(M^*_\delta\right) = N^*_\delta\,  \mathcal{I}_{\theta}\left(k^*_\delta+1,N^*_\delta-k^*_\delta\right)-\frac{k^*_\delta+1}{\theta}\mathcal{I}_{\theta}\left(k^*_\delta+2,N^*_\delta-k^*_\delta\right),
\ee
Next note that for sufficiently small $\delta$ (as $\delta\to 0$), we have by Lemma \ref{lem1} and \eqref{2.19}
$$
\mathcal{I}_{\theta}\left(k^*_\delta+2,N^*_\delta-k^*_\delta\right)\approx \mathcal{I}_{\theta}(k^*_\delta+1,N^*_\delta-k^*_\delta)=\Pi_{\operatorname{T^*_{seq}}}(\theta),
$$
and that, 
$$
\frac{k^*_\delta+1}{N^*_\delta}\rightarrow \theta_0, 
$$
which when combined in \eqref{3.36} together with the fact that $E_\theta\left(M^*_\delta\right) \leq N^*_\delta$, provides that, 
$$
E_\theta\left(\frac{N^*_\delta-M^*_\delta}{N^*_\delta}\right) =\left(1-\frac{\theta_0}{\theta}\right)^+\Pi_{\operatorname{T^*_{seq}}}(\theta), 
$$ 
as $\delta \rightarrow 0$. Finally, \eqref{3.35} follows from Lemma \ref{lem2} below, noting that as $\delta\to 0$,
$$
\Pi_{\operatorname{T^*_{seq}}}(\theta)\to \begin{cases} & 0, \qquad \theta<\theta_0; \\
    & \alpha,  \qquad  \theta=\theta_0;\\
    & 1,  \qquad \theta>\theta_0.
\end{cases}
$$
\end{prof}

In the Figure \ref{f3} below, we show that as $\delta$ decreasing, the relative 'savings' in the ASN converges to the asymptotic one which supports Theorem \ref{thrm3}.

\begin{figure}[H]
\centering
\begin{subfigure}[b]{0.327\textwidth}
\centering
\includegraphics[width=\textwidth]{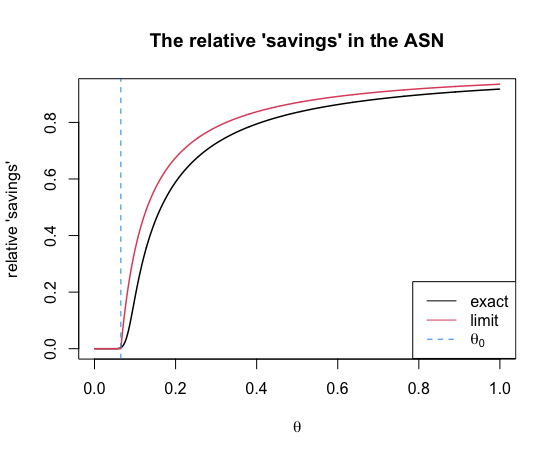}
\caption{$\delta=0.5$}
\end{subfigure}
\hfill
\begin{subfigure}[b]{0.327\textwidth}
\centering
\includegraphics[width=\textwidth]{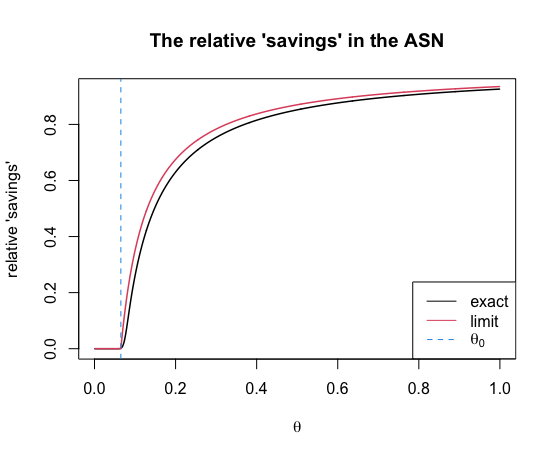}
\caption{$\delta=0.25$}
\end{subfigure}
\begin{subfigure}[b]{0.327\textwidth}
\centering
\includegraphics[width=\textwidth]{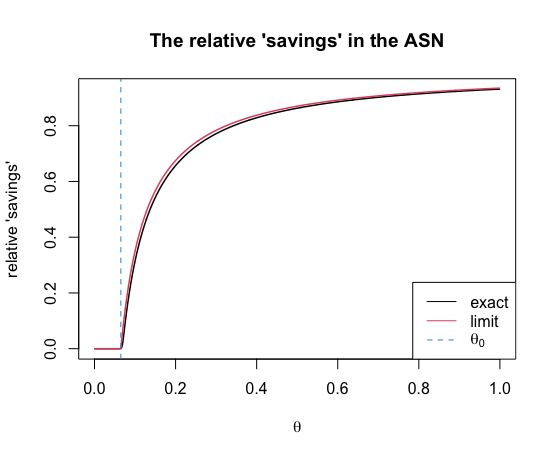}
\caption{$\delta=0.1$}
\end{subfigure}
\caption{The plots of the relative 'savings' in the ASN with respect to $\alpha=0.05$, $\beta=0.1$, $\theta_0=0.065$ and different values of $\delta$: (a) $\delta=0.5$: $\theta_1=0.0975$, $N^*=584$ and $k^*=47$; (b) $\delta=0.25$, $\theta_1=0.08125$, $N^*=2162$ and $k^*=159$; (c) $\delta=0.1$, $\theta_1=0.0715$, $N^*=12811$ and $k^*=878$.}
\label{f3}
\end{figure}

In the following theorem we provide the asymptotic mean and variance of the post-test estimator, $\hat{\theta}_{M^*_\delta}$, of $\theta$, namely of the rate of side effect incidence. 
\begin{thm}  \label{thrm4}
Let $\hat{\theta}_{M^*_\delta}$ be the post-test estimator (in \eqref{3.26}) of $\theta$ based on the $M^*_\delta=\min \left\{N^*_\delta, M_\delta\right\}$ observations obtained from the optimal $(\alpha, \beta)$ sequential test with $N_\delta^*$ and $k_\delta^*$ as are defined in \eqref{3.33} and \eqref{3.34} above. Then as $\delta\to 0$, we have for all $\theta \in (0,1)$, 
\be \label{3.37}
E_{\theta}\left(\hat{\theta}_{M^*_\delta}\right) \rightarrow \theta, \qquad \text{and} \qquad Var_{\theta}\left(\hat{\theta}_{M^*_\delta}\right) \rightarrow 0, 
\ee
and hence,
$\hat{\theta}_{M^*_\delta}\xrightarrow{P} \theta$.
\end{thm}
\begin{prof} 
See Section \ref{s6} below.
\end{prof}

In Table \ref{t3} below we present the calculated mean and variance of $\hat{\theta}_{M^*}$ as are calculated for various values of $\delta \ (0.1, 0.05, 0.01)$. Again, as can be seen, with $\delta \to 0$, $E_{\theta}(\hat{\theta}_{M^*_\delta})$ is close to the true $\theta$ and $Var_{\theta}(\hat{\theta}_{M^*_\delta})$ is close to $0$, as is expected from Theorem \ref{thrm4}.

\begin{table}[H]
\begin{center}{\small 
\caption{The expectation and the variance of $\hat{\theta}_{M^*}$ for fixed $\alpha=0.05$, $\beta=0.1$, $\theta_0=0.065$, with (i) $\delta=0.1 \ (N^*_\delta=12811\text{ and } k^*_\delta=878)$; (ii) $\delta=0.05 \ (N^*_\delta=50269 \text{ and } k^*_\delta=3358)$ and (iii) $\delta=0.01 \ (N^*_\delta=1236886\text{ and } k^*_\delta=80848)$.}
\label{t3}
\begin{tabular}{c|c|c|c|c|c|c}
\hline
&\multicolumn{3}{c}{$E_{\theta}\left(\hat{\theta}_{M^*_\delta}\right)$} \vline & \multicolumn{3}{c}{$Var_{\theta}\left(\hat{\theta}_{M^*_\delta}\right)$} \\
\cline{2-4}\cline{5-7}
$\theta$& $\delta=0.1$  & $\delta=0.05$ & $\delta=0.01$ & $\delta=0.1$  & $\delta=0.05$ & $\delta=0.01$ \\
\hline
0.065 & 0.064742 & 0.064875 & 0.064975  & 2.0804e-05 & 9.1591e-06 & 1.6428e-06   \\
0.1& 0.100103  & 0.100027 & 0.100001 & 1.0279e-05 & 2.6821e-06 &1.1132e-07   \\
0.2& 0.200182 &  0.200048 & 0.200002  & 3.6521e-05 & 9.5346e-06 & 3.9581e-07   \\
0.3& 0.300239  & 0.300063 & 0.300003 & 7.1852e-05 & 1.8768e-05 & 7.7925e-07   \\
0.4& 0.400273 & 0.400072 & 0.400003 & 1.0941e-04 & 2.8594e-05 & 1.1874e-06   \\
0.5& 0.500284 & 0.500074 & 0.500003 & 1.4237e-04 & 3.7225e-05 & 1.5461e-06   \\
\hline
\end{tabular}
}
\end{center}
\end{table} 

In the next theorem we present the main results of this paper, establishing the asymptotic normality of the post-test estimator, $\hat{\theta}_{M^*_\delta}$, appropriately standardized.

\begin{thm}  \label{thrm5}	
$\forall$ $\theta\in (0,1)$, as $\delta\rightarrow 0$, we have
\be \label{3.38}
{\cal U}_\delta:=\sqrt{M^*_\delta}(\hat{\theta}_{M^*_\delta}-\theta)\xrightarrow{\text{D}} {\N}(0,\theta(1-\theta)). 
\ee
\end{thm}

\begin{prof}
Note at first that since $M_{\delta}\equiv M_{k_\delta^*}$ and $M^*_\delta\equiv \min\{M_{\delta}, N^*_\delta \}$, we have for any $t \in \Bbb{R}$, 
$$
\operatorname{P}_\theta \left({\cal U}_\delta \leq t \right)= \operatorname{P}_\theta \left({\cal U}_\delta \leq t\,  \mid  M_\delta > N^*_\delta \right)\operatorname{P}_\theta \left(M_\delta > N^*_\delta \right)+ \operatorname{P}_\theta \left({\cal U}_\delta \leq t\,  \mid  M_\delta\leq N^*_\delta \right)\operatorname{P}_\theta \left(M_\delta\leq N^*_\delta \right). 
$$
Hence, 
\begin{multline*} 
\lim_{\delta \rightarrow 0}\operatorname{P}_\theta \left( {\cal U}_\delta \leq t \right)=\lim_{\delta \rightarrow 0}\operatorname{P}_{\theta}\left(\sqrt{N^*_\delta}\left(\frac{S_{N^*_{\delta}} }{N^*_{\delta}}-\theta \right)\le t  \right)\cdot \lim_{\delta \rightarrow 0}\left[1-\Pi_{\operatorname{T^*_{seq}}}(\theta)\right] \\
+\lim_{\delta \rightarrow 0}\operatorname{P}_{\theta}\left(\sqrt{M_\delta}\left(\frac{k^*_{\delta}+1}{M_\delta}-\theta\right) \le t \right)\cdot\lim_{\delta \rightarrow 0}\Pi_{\operatorname{T^*_{seq}}}(\theta).  
\end{multline*}
Obviously, since $S_{N^*_{\delta}} \sim \mathcal{B}(N^*_\delta, \theta)$, by CLT, we obtain the asymptotic normality that
\be \label{3.39}
\sqrt{N^*_{\delta}}\left(\frac{S_{N^*_{\delta}} }{N^*_{\delta}}-\theta\right)\stackrel{D}{\rightarrow} \N\left(0,\theta(1-\theta)\right), \quad \text{as} \quad \delta \rightarrow 0.
\ee
Following Lemma \ref{lem3} below (which is a restatement of Rényi's adaptation of Anscombe’s theorem, see Section \ref{s6}), we also obtain that 
\be \label{3.40}
\sqrt{M_\delta}\left(\frac{k^*_{\delta}+1}{M_\delta}-\theta \right) \xrightarrow{D} \N \left(0,\theta(1-\theta)\right), \quad \text { as } \quad {\delta} \rightarrow 0.
\ee
The detailed proof (\ref{3.40}) are provided in Section \ref{s6}. Hence, upon combining \eqref{3.39} and \eqref{3.40} together, we obtain
\begin{align*} 
\lim_{\delta \rightarrow 0}\operatorname{P}_\theta \left( {\cal U}_\delta \leq t \right) 
&=\Phi \left(\frac{t}{\sqrt{\theta(1-\theta)}}\right)\cdot \lim_{\delta \rightarrow 0}\left[1-\Pi_{\operatorname{T^*_{seq}}}(\theta)\right] +\Phi \left(\frac{t}{\sqrt{\theta(1-\theta)}}\right)\cdot\lim_{\delta \rightarrow 0}\Pi_{\operatorname{T^*_{seq}}}(\theta)\notag \\
&=\Phi \left(\frac{t}{\sqrt{\theta(1-\theta)}}\right),
\end{align*}
which indicates that
$$
\sqrt{M^*_\delta}\left(\hat{\theta}_{M^*_\delta}-\theta \right) \xrightarrow{D} \N \left(0,\theta(1-\theta)\right), \quad \text { as } \quad {\delta} \rightarrow 0,
$$
as stated in \eqref{3.38}.
\end{prof}

In closing of this section we note that according to Theorems \ref{thrm4} and \ref{thrm5}, we may construct an approximate $(1-\gamma)\times 100\%$ confidence interval for $\theta$, when $N^*$ and $k^*$ are sufficiently large (when $\delta$ is sufficiently small), as,
\be \label{3.41}
\hat{\theta}_{M^*}\pm Z_{ \gamma/2}\sqrt{\frac{\hat{\theta}_{M^*}\left(1-\hat{\theta}_{M^*}\right)}{M^*}}.
\ee

To illustrate \eqref{3.41}, we performed a $10000$ simulations of a $95\%$ confidence interval for $\theta$. We set up the experiments with $\theta_0=0.065$, $\delta=0.2(0.1)$, $\theta_1=\theta_0(1+\delta)=0.078(0.0715)$, $\alpha=0.05$ and $\beta=0.1$. The results are presented in Table \ref{t4} below. 
\begin{table}[H]
\centering
\caption{The simulated coverage probability of $95\%$ confidence interval}
\begin{tabular}{c|cc|cc}
\hline
&\multicolumn{4}{c}{simulated coverage probability of $95\%$ CI}\\
\cline{2-5}
&\multicolumn{2}{c}{$\delta=0.2(\theta_1=0.078)$}\vline &\multicolumn{2}{c}{$\delta=0.1(\theta_1=0.0715)$} \\
\cline{2-3}\cline{4-5}
$\theta$&{$N^*=3321$}  & {$k^*=239$} & {$N^*=12811$} & {$k^*=878$} \\
\hline
0.05&\multicolumn{2}{c}{$94.55\%$}\vline &\multicolumn{2}{c}{$94.85 \%$} \\
0.065&\multicolumn{2}{c}{$94.68\%$}\vline &\multicolumn{2}{c}{$94.93\%$} \\
0.08&\multicolumn{2}{c}{$94.85\%$}\vline &\multicolumn{2}{c}{$95.22\%$} \\
0.2&\multicolumn{2}{c}{$95.01\%$}\vline &\multicolumn{2}{c}{$94.95\%$} \\
\hline
\end{tabular}
\label{t4}
\end{table}
As can be seen, all of the 'observed' coverage probabilities are close, as expected from Theorem \ref{thrm4}, to the nominal confidence level of $95\%$.

\section{Analysis of Some COVID-19 Side Effect Data}
\label{s4}

\citet{beatty2021analysis} provides a detailed analysis of adverse effects following COVID-19 vaccination during the period, March 26, 2020, to May 19, 2021. We utilize these data to illustrate our testing and estimation procedure. To that end, we focus attention on reported side effect, namely, allergic reaction (or anaphylaxis). In all, there were $53$ reported cases of allergic reaction among the $19821$ participants during that period, as recorded after first or second doses of the vaccine (the BNT162b2 of Pfizer/BioNTech, the mRNA-1273 of Moderna, or the JNJ-78436735, of Johnson $\&$ Johnson). We view the conclusion of the reported observation period as the termination of a sequential testing procedure of $H_0$ as given in \eqref{1.1}, with the two possible outcomes of the following scenarios in which: $(i)$ $H_0$ is not rejected and $(ii)$ $H_0$ is being rejected. The following figures illustrate our proposed sequential test procedure in these two scenarios.

\begin{figure}[H]
\centering
\begin{subfigure}[b]{0.48\textwidth}
\centering
\begin{tikzpicture}[>=Latex,scale=0.2]
\draw[help lines, color=white](0,0) grid (25,8);
\fill (0,0) node[left]{$O$} circle (.1);
\draw[thick,->] (0,0) -- (27,0) node[anchor=north west]{$n$};
\draw[thick,->] (0,0) -- (0,10) node[anchor=south west]{$S_n$};
\fill (25,0) node[below]{\small $19821$} circle (.2);
\draw[semithick, dashed](25,0)--(25,8);
\draw (25,4) node[right]{\small $(19821,53)$} circle (.2);
\fill (0,8) node[left]{\small $116$} circle (.2);
\fill (8,0) node[below]{\small $116$}circle (.2) ;
\draw[semithick,solid](8,8)--(25,8);
\draw[semithick,dashed,color=gray!60](0,4)--(25,4);
\fill (8,8)  circle (.2);
\fill (25,8)  circle (.2);
\draw[semithick, dashed] (0,0) -- (8,8);
\fill (0,4) node[left]{\small $53$}circle (.2) ;
\draw[semithick,solid] (0,0) -- (2,0)--(4,1)--(8,1)-- (10,2)-- (13,2)-- (17,3)--(20,3)--(23,4)--(25,4);
\fill [brown](25,4)  circle (.2);
\end{tikzpicture}
\caption{scenario $(i)$ (non-rejection) }
\end{subfigure}
\hfill
\begin{subfigure}[b]{0.48\textwidth}
\centering
\begin{tikzpicture}[>=Latex,scale=0.2]
\draw[help lines, color=white](0,0) grid (25,8);
\fill (0,0) node[left]{$O$} circle (.1);
\draw[thick,->] (0,0) -- (27,0) node[anchor=north west]{$n$};
\draw[thick,->] (0,0) -- (0,10) node[anchor=south west]{$S_n$};
\draw[semithick, dashed,color=gray!60](24,0)--(24,8);
\fill (25,0) circle (.2);
\fill (25,1) node[right]{\small $20934$};
\fill (24,0) circle (.2);
\fill (23,0) node[below]{\small $19821$};
\draw[semithick, dashed](25,0)--(25,8);
\fill (24,8) node[above]{\small $(19821,53)$} circle (.2);
\fill (0,8) node[left]{\small $53$} circle (.2);
\draw[semithick,solid](8,8)--(25,8);
\fill (8,8)  circle (.2);
\fill (25,8)  circle (.2);
\draw[semithick, dashed] (0,0) -- (8,8);
\fill (8,0) node[below]{\small $53$}circle (.2) ;
\draw[semithick,solid] (0,0) -- (1,0)-- (3,1)--(5,1)--(7,2)-- (10,2)--(13,3)--(15,3)--(18,4)--(20,6)--(21,6)--(23,7)-- (24,8);
\end{tikzpicture}
\caption{scenario $(ii)$ (rejection)}
\end{subfigure}
\caption{The illustrations of the curtailed sequential test $\operatorname{T^*_{seq}}$ based on the \citet{beatty2021analysis} data for two scenarios}

\end{figure}

$\bullet$\  \underbar{In scenario $(i)$} ({\it{a non-rejection of $H_0$}}) Here we assume $\theta_0=0.005$, then the pair of the test hypotheses are:
$$
H_0: \theta \le 0.005 \ \ \text{against} \ \ H_1: \theta > 0.005.
$$
According to \eqref{2.13}, $M_{k^*}>N^*\equiv 19821$, $M^*=\min\{M_{k^*},N^*\}=19821$ and $S_{N^*}=53$. 
With $\alpha=0.05$, we use \eqref{2.9}, to calculate the critical value $k^*=115$. Utilizing \eqref{2.8}, to let $\beta$ close to $0.10$, we assume $\theta_1=0.0065$. We calculate the corresponding $\tilde{\alpha}$ and $\tilde{\beta}$ of this test by \eqref{2.6} and \eqref{2.7}, that is,
$$
\tilde{\alpha}=\operatorname{P}_{\theta_0}(S_{N^*} >k^*)=0.0494 \quad \text{and} \quad \tilde{\beta}=\operatorname{P}_{\theta_1}(S_{N^*} \le k^*)=0.1192.
$$
Therefore, for this optimal $(\alpha,\beta)$ sequential test with $\alpha \approx 0.05$ and $\beta \approx 0.12$, with total sample size $N^* \equiv 19821$ and the critical value $k^*=115$, we would not reject the null hypothesis that the probability of allergic reaction or anaphylaxis $\theta$ is less than or equal to $0.005$.

\vskip 15pt

$\bullet$\  \underbar{In scenario $(ii)$} ({\it{a rejection of $H_0$}}) Here we assume $\theta_0=0.002$, then the pair of the test hypotheses are:
$$
H_0: \theta \le 0.002 \ \ \text{against} \ \ H_1: \theta > 0.002.
$$ 
According to \eqref{2.13}, $M_{k^*}\equiv 19821 < N^*$, $M^*=\min\{M_{k^*},N^*\}=19821$ with $k^*=52$ and $S_{M_{k^*}}=53$. With $\alpha=0.05$, we solve equation \eqref{2.9}, to obtain that the maximal sample size $N^*=20934$. Utilizing \eqref{2.8}, to let $\beta$ close to $0.1$, we assume $\theta_1=0.003$. We calculate the corresponding $\tilde{\alpha}$ and $\tilde{\beta}$ of this test by \eqref{2.6} and \eqref{2.7}, that is,
$$
\tilde{\alpha}=\operatorname{P}_{\theta_0}(S_{N^*} >k^*)=0.0500 \quad \text{and} \quad \tilde{\beta}=\operatorname{P}_{\theta_1}(S_{N^*} \le k^*)=0.0965.
$$
Therefore, for this optimal $(\alpha,\beta)$ sequential test with $\alpha \approx 0.05$ and $\beta \approx 0.10$, with total sample size $N^*=20934$ and the critical value $k^* \equiv 52$, we would stop the study and reject the null hypothesis.

Note that in both scenarios, $M^*=19821$ and $S_{M^*}=53$, and therefore, our post-test (terminal) estimate of $\theta$ is $\hat{\theta}$ as given by 
$$
\hat{\theta}=\frac{S_{M^*}}{M^*}=\frac{53}{19821}=0.0027.
$$
Now, by applying \eqref{3.41} and substituting the values of $\hat{\theta}$ and $M^*$, we obtain that the $95\%$ confidence interval of the true probability of allergic reaction or anaphylaxis, $\theta$, as
$$
0.001955\ \leq \theta\leq \ 0.003393.
$$

\section{Summary and Discussion}\label{s5}

In this paper we construct an $(\alpha, \beta)$-{\it{optimal}} (non-randomized) sequential testing procedure for early detection of treatment's side effect. While we cast the problem in context of a vaccination campaign (i.e. COVID-19 immunization), our results are equally applicable to other 'treatments' scenario that may results with undesired side effect(s), which their early detection is paramount. The novelty of our results is based on an integral representation of binomial probabilities, making repeated use of regularized incomplete beta function. This representation is key for demonstrating the optimality of our testing procedure as well as of many of the calculations involved.

We express the sequential test in terms of the 'stopping time' involved (i.e. the random sample size) and provide explicit expressions to its mean (the so-called ASN function) and its variance. The asymptotic properties of the stopping time are also provided under the local (contiguous) alternative case. Additionally, we propose a natural post-test estimator of the unknown proportion of the side effect and investigate its properties. Unlike the MLE based estimator (see \citet{phatak1967estimation}) which hinges on the asymptotic normality of the MLE in general, we establish and prove the consistency and the asymptotic normality of our post-test (randomly stopped) estimator of the side effect's probability.

These results are instrumental in providing the appropriate confidence interval of the unknown parameter, $\theta$, the probability of a side effect. With our approach, we are able to provide exact calculation of the variance of the post-test estimator, as well as the coverage probabilities. We point out that, in contrast, any asymptotic MLE-base variance calculation (e.g. \citet{phatak1967estimation}) will lead to underestimation of the variance when $\theta$ is small and to variance overestimation when $\theta$ is large, hence impacting the desired accuracy of the confidence interval.

As can be seen from the COVID-19 data analysis of Section \ref{s4}, our procedure performs well in the most realistic scenarios depicted in the Example of a non-rejection and of a rejection of the null-hypotheses leading to a reliable and intuitively appealing post-test estimate of the unknown proportion, while also  meeting the desired levels of the test's error probabilities. It should be clear however, that our proposed testing procedure is particularly useful when applied in context of {\it large populations}, such as in the case of the COVID-19 vaccination campaign, which allows for the large-sample asymptotic behaviour to prevail.

This is especially notable in post-marketing surveillance data, for the EUA treatment safety analysis (of the kind employed for the COVID-19), our proposed sequential test may be a good choice to apply (dealing inherently with large impacted populations). It would provide an efficient way to protect participants, through an early detection of potential unknown side effect(s). In a subsequent work, we discuss how sequential procedure can be utilized for the early detection of two (or multiple) possible side effects of a given treatment. Clearly, the detailed analysis in this paper of the sequential test for a single side effect, plays an important role in our subsequent work.

\section{Some Additional Technical Results and Proofs}\label{s6}
In this section we provide some auxiliary technical results and the proofs of Lemma \ref{lem1} and \ref{lem2} and Theorem \ref{thrm4} and of \eqref{3.40} in Theorem \ref{thrm5}. We begin with the proof of Lemma \ref{lem1}.

\begin{prof} \hskip  -10pt of \ul{Lemma 1}: \ Since by \eqref{3.33},
$$
\lim_{\delta \rightarrow 0}
N^*_\delta=\lim_{\delta \rightarrow 0}\left[\frac{z_{\alpha}}{\delta}\sqrt{\frac{1-\theta_0}{\theta_0}}+z_{\beta}\sqrt{\frac{1+\delta}{\delta^2}\left(\frac{1}{\theta_0}-1-\delta\right)}\right]^2 = \infty.
$$
Also, by the expression \eqref{3.34} of $k^*_\delta$, we  have
$$
\lim_{\delta \rightarrow 0}k^*_\delta=\lim_{\delta \rightarrow 0}N^*_\delta \left[ z_\alpha\sqrt{\frac{\theta_0(1-\theta_0)}{N^*_\delta}}+\theta_0 \right]-\frac{1}{2}= \infty. 
$$
Accordingly, it follows immediately that, 
\begin{align*} 
\lim_{\delta \rightarrow 0} \frac{k^*_\delta}{N^*_\delta}&=\lim_{\delta \rightarrow 0}\frac{N^*_\delta \left[z_\alpha\sqrt{\frac{\theta_0(1-\theta_0)}{N^*_\delta}}+\theta_0 \right]-\frac{1}{2}}{N^*_\delta} \nonumber\\
&=\lim_{\delta \rightarrow 0}\left(z_\alpha\sqrt{\frac{\theta_0(1-\theta_0)}{N^*_\delta}}+\theta_0-\frac{1}{2N^*_\delta} \right) \nonumber\\
&= \theta_0,
\end{align*}
which completes the proof of Lemma \ref{lem1}. 
\end{prof}

In the next lemma, we establish the limiting value of the power function $\Pi_{\operatorname{T^*_{seq}}}(\theta)$ as $\delta\to 0$ (see also the proof of Theorem \ref{thrm3} and \eqref{3.35}). 

\begin{lemma} \label{lem2}
As $\delta \rightarrow 0$, the power function in our proposed curtailed sequential test, $\Pi_{\operatorname{T^*_{seq}}}(\theta)$, is given by, 
\be \label{5.42}
\lim_{\delta \rightarrow 0}\Pi_{\operatorname{T^*_{seq}}}(\theta)=
\begin{cases}
	0, \  \ & \text{if } \theta <\theta_0;\\
	\alpha, \  \ & \text{if } \theta =\theta_0; \\
	1, \ \ & \text{if } \theta> \theta_0.
\end{cases}
\ee
\end{lemma}
\begin{prof}
By Theorem \ref{thrm1}, the properties of $\Pi_{\operatorname{T^*_{seq}}}(\theta)$ are equivalent to those properties of $\Pi_{\operatorname{T_{fix}}}(\theta)$. Therefore, we may consider the asymptotic behavior of $\Pi_{\operatorname{T_{fix}}}(\theta)$, by applying the Normal Approximation to the underlying binomial assumption that with a fixed $N^*_\delta$, $S_{N^*_\delta} \sim \mathcal{B}(N^*_\delta, \theta)$. Accordingly we have,
\begin{align*}  
 \lim_{\delta \rightarrow 0}\Pi_{\operatorname{T_{fix}}}(\theta)&=\lim_{\delta \rightarrow 0}\operatorname{P}_{\theta}\left(S_{N^*_\delta} > k^*_\delta\right) \notag\\
&=1-\lim_{\delta \rightarrow 0}\Phi\left( \frac{k^*_\delta+\frac{1}{2}-N^*_{\delta}\theta}{\sqrt{N^*_{\delta}\theta(1-\theta)}} \right) \\
&= 1-\lim_{\delta \rightarrow 0}\Phi\left( \frac{z_{\alpha}\sqrt{N^*_{\delta}\theta_0(1-\theta_0)}+N^*_{\delta}\theta_0-N^*_{\delta}\theta}{\sqrt{N^*_{\delta}\theta(1-\theta)}} \right) \\
&:= 1-\lim_{\delta \rightarrow 0}\Phi(A_\delta),
\end{align*}
where 
$$
A_\delta:= \frac{z_{\alpha}\sqrt{\theta_0(1-\theta_0)}+\sqrt{N^*_{\delta}}(\theta_0-\theta)}{\sqrt{\theta(1-\theta)}}. 
$$
But,  
\begin{itemize}
\item  If $\theta<\theta_0$, it follows immediately from Lemma \ref{lem1} that $\lim_{\delta\to 0}A_{\delta}=+\infty$ and hence, 
$$
\lim_{\delta \rightarrow 0}\Pi_{\operatorname{T_{fix}}}(\theta) =0.
$$
\item If $\theta=\theta_0$, clearly, $A_{\delta}\equiv z_\alpha$ and hence, $\Pi_{\operatorname{T_{fix}}}(\theta_0)\equiv \alpha$. 

\item If $\theta>\theta_0$, it follows immediately from Lemma \ref{lem1} that $\lim_{\delta\to 0}A_{\delta}=-\infty$ and hence, 
$$
\lim_{\delta \rightarrow 0}\Pi_{\operatorname{T_{fix}}}(\theta) =1.
$$
\end{itemize}

Finally, by combining above three situations together we obtain \eqref{5.42}.
\end{prof}

\begin{prof}   \hskip  -10pt \ \textbf{\ul{of Theorem 4}:}

{Recall the expression \eqref{3.27} for $E(\hat \theta_{M^*_\delta})$ with its two parts \eqref{3.28} and \eqref{3.29}. Clearly by \eqref{3.28}, 
$$
\lim_{\delta \to 0}E_\theta\left(\frac{S_{N^*_\delta} }{N^*_\delta} \mathbbm{1}\left[S_{N^*_\delta}\le k^*_\delta\right]\right)=\theta \lim_{\delta \to 0}\operatorname{P}_{\theta}\left(S_{N^*_\delta-1} \le k^*_\delta-1\right).
$$
By \eqref{2.18} we have  
$$
\operatorname{P}_{\theta}\left(S_{N^*_\delta-1} \le k^*_\delta-1\right)=1-\mathcal{I}_{\theta}(k^*_\delta-1,N^*_\delta-k^*_\delta+1). 
$$
However, since by Lemma \ref{lem1}, as $\delta\to 0$, $k^*_\delta+1\approx k^*_\delta-1$, 
$$
\mathcal{I}_{\theta}(k^*_\delta-1,N^*_\delta-k^*_\delta+1) \approx \mathcal{I}_{\theta}(k^*_\delta+1,N^*_\delta-k^*_\delta)=\Pi_{\operatorname{T^*_{seq}}}(\theta).
$$
Hence, we have
\be \label{5.43}
\lim_{\delta \to 0}\operatorname{P}_{\theta}\left(S_{N^*_\delta-1} \le k^*_\delta-1\right)=1-\lim_{\delta \to 0}\Pi_{\operatorname{T^*_{seq}}}(\theta),
\ee
and
\be \label{5.44}
\lim_{\delta \to 0}E_\theta\left(\frac{S_{N^*_\delta} }{N^*_\delta} \mathbbm{1}\left[S_{N^*_\delta}\le k^*_\delta\right]\right)=\theta \left[1-\lim_{\delta \to 0}\Pi_{\operatorname{T^*_{seq}}}(\theta)\right].
\ee
Now, by \eqref{3.29},
$$
\lim_{\delta \to 0}E_{\theta}\left(\frac{k^*_\delta+1}{M_\delta}\mathbbm{1}\left[M_\delta \le N^*_\delta\right]\right)=(k^*_\delta+1)\lim_{\delta \to 0} \sum_{j=k^*_\delta+1}^{N^*_\delta}\frac{1}{j}\operatorname{P}_{\theta}(M_\delta=j),
$$
and since $M_\delta\to \infty$ ($M_\delta > k^*_\delta$), as $\delta\to \infty$, we may approximate $\frac{1}{M_\delta}$ as $\frac{1}{M_\delta-1}$ to obtain that
\begin{align} \label{5.45}
\lim_{\delta \to 0}E_{\theta}\left(\frac{k^*_\delta+1}{M_\delta}\mathbbm{1}\left[M_\delta \le N^*\right]\right)& = (k^*_\delta+1)\lim_{\delta \to 0} \sum_{j=k^*_\delta+1}^{N^*_\delta}\frac{1}{j-1}\operatorname{P}_{\theta}(M_\delta=j) \notag \\
&=\theta\lim_{\delta \rightarrow 0}\frac{ \left(k^*_{\delta}+1 \right)}{k^*_{\delta}}\operatorname{P}_\theta \left(\tilde{M} \le N^*_\delta-1\right) \notag\\
&=\theta\lim_{\delta \rightarrow 0}\frac{ \left(k^*_{\delta}+1 \right)}{k^*_{\delta}}\mathcal{I}_{\theta}\left(k^*_{\delta}, N^*_{\delta}-k^*_{\delta}\right) \notag\\
&=\theta\lim_{\delta \rightarrow 0}\mathcal{I}_{\theta}\left(k^*_{\delta}+1, N^*_{\delta}-k^*_{\delta}\right) \notag\\
&= \theta \lim_{\delta \rightarrow 0}\Pi_{\operatorname{T^*_{seq}}}(\theta),
\end{align}
where $\tilde{M}$ denotes a $\mathcal{NB}\left(k^*_\delta,\theta\right)$ random variable and by utilizing again \eqref{2.18} in the second equality. Accordingly, by combining \eqref{5.44} and \eqref{5.45} together, we obtain
$$
\lim_{\delta \to 0}E_{\theta}\left(\hat{\theta}_{M^*_\delta}\right)=\theta\left[1-\lim_{\delta \to 0}\Pi_{\operatorname{T^*_{seq}}}(\theta)\right]+\theta \lim_{\delta \rightarrow 0}\Pi_{\operatorname{T^*_{seq}}}(\theta)=\theta.
$$
Similarly, for the limiting value of $Var_{\theta}(\hat{\theta}_{M^*_\delta})$, we consider the limiting values of \eqref{3.31} and \eqref{3.32} for the second moment in \eqref{3.30}. Recall the expression in \eqref{3.31}, that
$$
\lim_{\delta \to 0}E_{\theta}\left(\frac{S_{N^*_\delta}^2 }{{N^*_\delta}^2}\mathbbm{1}\left[S_{N^*_\delta}\leq k^*_\delta\right]  \right)=\lim_{\delta \to 0}\frac{(N^*_\delta-1)\theta^2\operatorname{P}_{\theta}\left(S_{N^*_\delta-2}\le k^*_\delta-2\right)}{N^*_\delta}+\lim_{\delta \to 0}\frac{\theta\operatorname{P}_{\theta}\left(S_{N^*_\delta-1} \le k^*_\delta-1\right)}{N^*_\delta}.
$$
Now, since by Lemma \ref{lem1} and by \eqref{2.18}, we have
\begin{align}\label{5.46}
\lim_{\delta \to 0}\operatorname{P}_{\theta}\left(S_{N^*_\delta-2}\le k^*_\delta-2\right)&=1-\lim_{\delta \to 0}\mathcal{I}_{\theta}(k^*_\delta-2,N^*_\delta-k^*_\delta+1) \notag\\
&= 1-\lim_{\delta \to 0}\mathcal{I}_{\theta}(k^*_\delta+1,N^*_\delta-k^*_\delta) \notag \\
&=1-\lim_{\delta \to 0}\Pi_{\operatorname{T^*_{seq}}}(\theta).
\end{align}
Then by \eqref{5.43} and \eqref{5.46}, we obtain
\begin{align}\label{5.47}
\lim_{\delta \to 0}E_\theta\left(\frac{S_{N^*_\delta}^2}{{N^*_\delta}^2} \mathbbm{1}\left[S_{N^*_\delta}\le k^*_\delta\right]\right)&=\theta^2\left[1-\lim_{\delta \to 0}\frac{N^*_\delta-1}{N^*_\delta}\Pi_{\operatorname{T^*_{seq}}}(\theta)\right]+\lim_{\delta \to 0}\frac{\theta}{N^*_\delta}\left[1-\Pi_{\operatorname{T^*_{seq}}}(\theta)\right] \notag \\
&=\theta^2\left[1-\lim_{\delta \to 0}\Pi_{\operatorname{T^*_{seq}}}(\theta)\right].
\end{align}
Again, since $\lim_{\delta \rightarrow 0}M_\delta=\infty$, we approximate $\frac{1}{M_\delta^2}$ as $\frac{1}{(M_\delta-1)(M_\delta-2)}$ in \eqref{3.32} to obtain, 
\begin{align} \label{5.48}
\lim_{\delta \to 0}E_{\theta}\left(\frac{(k^*_\delta+1)^2}{M^2_\delta}\mathbbm{1}\left[M_\delta \le N^*_\delta \right] \right)&=\lim_{\delta \to 0}(k^*_\delta+1)^2 \sum_{j=k^*_\delta+1}^{N^*_\delta}\frac{1}{(j-1)(j-2)}\operatorname{P}_{\theta}(M_\delta=j) \notag\\
&=\theta^2\lim_{\delta \rightarrow 0}\frac{ (k^*_{\delta}+1)^2}{k^*_{\delta}(k^*_{\delta}-1)}\operatorname{P}_\theta \left(\tilde{M} \le N^*_\delta-2\right) \notag\\
&=\theta^2\lim_{\delta \rightarrow 0}\frac{ (k^*_{\delta}+1)^2}{k^*_{\delta}(k^*_{\delta}-1)}\mathcal{I}_{\theta}\left(k^*_{\delta}-1, N^*_{\delta}-k^*_{\delta} \right) \notag\\
&=\theta^2 \lim_{\delta \rightarrow 0}\mathcal{I}_{\theta}\left(k^*_{\delta}+1, N^*_{\delta}-k^*_{\delta} \right) \notag\\
&=\theta^2 \lim_{\delta \rightarrow 0}\Pi_{\operatorname{T^*_{seq}}}(\theta).
\end{align}
where $\tilde{M} \sim \mathcal{NB}\left(k^*_\delta-1,\theta\right)$ in the second equality and utilizing \eqref{2.18} in the second equality. Combining \eqref{5.47} and \eqref{5.48} together, we obtain
$$
\lim_{\delta \to 0} E_{\theta}\left(\hat{\theta}^2_{M^*_\delta}\right)=\theta^2 \left[1-\lim_{\delta \rightarrow 0}\Pi_{\operatorname{T^*_{seq}}}(\theta)\right]+\theta^2 \lim_{\delta \rightarrow 0}\Pi_{\operatorname{T^*_{seq}}}(\theta)=\theta^2.
$$
Accordingly, since $\lim_{\delta \to 0}E_{\theta}(\hat{\theta}_{M^*_\delta})=\theta$, we obtain that
$$
\lim_{\delta \to 0}Var_{\theta}\left(\hat{\theta}_{M^*_\delta}\right)=0,
$$
which completes the proof of \eqref{3.37}. Finally we note that \eqref{3.37} implies that $\hat{\theta}_{M^*_\delta}\xrightarrow{\text{P}} \theta$, $\forall \ \theta \in (0,1)$.
}
\end{prof}

The following lemma is critical in establishing the result in \eqref{3.40} concerning the asymptotic behavior of stopping time $M^*_\delta$. It is based on Rényi's adaptation of Anscombe’s theorem (see \citet{renyi1960central}) and is restated (see Theorem $2.3$ in \citet{gut2012anscombe}) here without a proof. 

\begin{lemma}[Rényi's adaptation of Anscombe’s theorem] \label{lem3}
Let $X_1$,$X_2$,$\ldots$ be i.i.d. random variables with mean 0 and positive, finite, variance $\sigma^2$, set $S_n=\sum_{k=1}^n X_k, n \geq 1$, and suppose that $\{\tau(t), t \geq 0\}$ is a family of positive, integer valued random variables, such that
$$
\frac{\tau(t)}{t} \stackrel{P}{\rightarrow} \eta \quad(0<\eta<\infty) \quad \text { as } \quad t \rightarrow \infty .
$$
Then
$$
\frac{S_{\tau(t)}}{\sigma \sqrt{\tau(t)}} \stackrel{D}{\rightarrow} \mathcal{N}(0,1) \quad \text { and } \quad \frac{S_{\tau(t)}}{\sigma \sqrt{\eta t}} \stackrel{D}{\rightarrow} \mathcal{N}(0,1) \quad \text { as } \quad t \rightarrow \infty.
$$
\end{lemma} 

\begin{prof} \hskip  -10pt of  \ul{Equation $(40)$ in Theorem $5$:} 

By utilizing Lemma \ref{lem3} above in our case, assuming $\xi_i=X_i-\theta$, we have $E(\xi_i)=0$ and $Var(\xi_i)=\theta(1-\theta)$. 
Since $\forall$ $\epsilon>0$, by Chebyshev’s Inequality,
$$
\lim_{\delta \rightarrow 0}\operatorname{Pr}\left( \lvert \frac{M_\delta}{k^*_{\delta}+1} -\frac{1}{\theta} \lvert > \epsilon \right) \le \lim_{\delta \rightarrow 0} \frac{Var\left(M_\delta\right)}{(k^*_{\delta}+1)^2\epsilon^2}= \lim_{\delta \rightarrow 0} \frac{\frac{k^*_{\delta}+1}{\theta^2}}{(k^*_{\delta}+1)^2\epsilon^2}= 0,
$$
we have
$$
\frac{M_\delta}{k^*_{\delta}+1} \xrightarrow{P}\frac{1}{\theta} \quad \left(0<\frac{1}{\theta}<\infty \right) \quad \text { as } \quad {\delta} \rightarrow 0.
$$
Then applying Lemma \ref{lem3} (Rényi's adaptation of Anscombe’s theorem), since
$$ S_{M_\delta}=\sum_{i=1}^{M_\delta}\xi_i=(k^*_{\delta}+1)(1-\theta)+(M_\delta-k^*_{\delta}-1)(-\theta)=k^*_{\delta}+1-M_\delta \theta,
$$
we have
$$
\frac{k^*_{\delta}+1-M_\delta \theta}{\sqrt{\theta(1-\theta)M_\delta}}\xrightarrow{D} \mathcal{N}(0,1), \quad \text { as } \quad {\delta} \rightarrow 0,
$$
by transformation,
$$
M_\delta\frac{\frac{k^*_{\delta}+1}{M_\delta}-\theta}{\sqrt{\theta(1-\theta)M_\delta}}\xrightarrow{D} \mathcal{N}(0,1), \quad \text { as } \quad {\delta} \rightarrow 0,
$$
which establishes \eqref{3.40},
$$
\sqrt{M_\delta}\left(\frac{k^*_{\delta}+1}{M_\delta}-\theta \right) \xrightarrow{D} \mathcal{N} \left(0,\theta(1-\theta)\right), \quad \text { as } \quad {\delta} \rightarrow 0, 
$$
as required, to complete the Proof of Theorem \ref{thrm5}.
\end{prof}

\printbibliography

\end{document}